\documentclass[11pt]{article}

\begin{document}

\title{ Stern-Gerlach Experiment and Bohm Limit }

\author{Marco Frasca\footnote{e-mail:marcofrasca@mclink.it}\\
Via Erasmo Gattamelata, 3 \\
00176 Roma (Italy)}

\date{\today}


\maketitle

\begin{abstract}
Stern-Gerlach experiment is a paradigm of measurement theory in quantum mechanics. Notwithstanding
several analysis given in literature, no clear understanding of the apparent collapse has been given
so far. Indeed, one can imagine a Stern-Gerlach device where all environmental effects are removed and
ask the question on how the measurement goes on. In this letter we will prove that a the Stern-Gerlach
device behaves as a true measurement apparatus, as expected by the Copenaghen interpretation, by
the Ehrenfest theorem. In this way we recover, by other means, a limit on the Stern-Gerlach
device, given by Bohm, based on scrambling of phases due to a large oscillation frequency.
\end{abstract}

A Stern-Gerlach experiment represents a classical paradigm of quantum measurement. As it
can be seen by the classical description due to D. Bohm \cite{bohm}, we can see the appearance
of all the peculiarities of the measurement process. Indeed, an entangled state 
between the apparatus and the quantum system appears and it is very difficult to understand
what could make the ``collapse'' happens.

Recently, a proposal for the experimental verification of the reality of ``collapse'' has been
given in \cite{pen1}. Such an experiment is aimed to prove the physical reality of the ``collapse'' 
that some authors ascribe to some stochastic effects, maybe due to
quantum gravity \cite{pen2,grw,gjm,perc}. Besides, a theory
has been proposed where time is treated as a stochastic variable \cite{boni} and could also
explain the appearance of the final result of the Stern-Gerlach experiment.
Decoherence studies have been carried out \cite{dec1,dec2} and also a
semiclassical analysis \cite{semi}. But, if all the decoherence effects are removed, does a
Stern-Gerlach device still work? If the answer is yes, could we attribute the effect to some
spontaneous collapse as some theories seem to accreditate? 

The main aim of this letter is to answer to such questions. We will show that the behavior of
a Stern-Gerlach device can be completely understood by the Ehrenfest theorem and there is no
need to assume that some wave function collapse happens somewhere. Our result permits to
obtain the limit of working of a Stern-Gerlach device already obtained by Bohm through the
concept of scrambling of the phase due to a large oscillation frequency \cite{bohm}. We
just note that this concept is the same as that of a singular limit put forward by Berry to understand
the quantum-classical transition \cite{berry} and in \cite{fra1,fra2,fra3,fra4} assuming
that the thermodynamic limit should grant such a transition.

The relevance of this result relies on the fact that quantum mechanics can give a definite
answer to the main questions of this letter, putting the matter on a strict experimental
ground. Once again, decoherence could appear as an intrinsic effect of the quantum evolution
without the need to resort to toy models or metaphysical hypothesis.

A general Hamiltonian for the Stern-Gerlach experiment, that applies to spin-${1\over 2}$ but can
be straightforwardly extended to any spin, is \cite{dec1,dec2,semi}
\begin{equation}
    H = \frac{p^2}{2m} + \lambda\sigma_z + \epsilon x\sigma_z
\end{equation}
with $x$ and $p$ position and momentum of the particle, $\lambda$ and $\epsilon$ the
first terms of the series of the product $\mu H(x)$ being $\mu$ the magnetic moment of the
atom and $H(x)$ the inhomogeneous magnetic field. This approximation is known to be the proper
one in this case \cite{bohm}.

This Hamiltonian has two properties that make the problem simple. Firstly, it is exactly
solvable. Secondly, with respect to the Ehrenfest Theorem, the average values of position
and momentum follow the Hamilton equations \cite{mess}. The latter property cannot grant that
an atom in the Stern-Gerlach device does behave classically.

The unitary evolution operator is straightforward to write down and is ($\hbar=1$)
\begin{equation}
     U_0(t) = e^{-i\left(\frac{p^2}{2m} + \lambda\sigma_z + \epsilon x\sigma_z\right)t}.
\end{equation}
If one take a well-localized particle with momentum $p_0$, centered on $x_0$ and initial spreading $\sigma$
\begin{equation}
     \psi(x,0) = \frac{1}{\sqrt{\sigma\sqrt{\pi}}}\exp\left[ip_0x-\frac{(x-x_0)^2}{2\sigma^2}\right]
\end{equation}
the unitary evolution gives, assuming an initial superposition state for the spin
$|\chi\rangle=a|\uparrow\rangle + b|\downarrow\rangle$, $|a|^2+|b|^2=1$ and $|\uparrow\rangle$,
$|\downarrow\rangle$ eigenstates of $\sigma_z$, gives
\begin{equation}
\label{eq:ent}
    \psi(x,t) = a\psi_\uparrow(x,t)|\uparrow\rangle + b\psi_\downarrow(x,t)|\downarrow\rangle
\end{equation}
being
\begin{equation}
    \psi_{\uparrow\downarrow}(x,t)=\sqrt{\frac{\sigma}{\sqrt{\pi}}}
	\frac{1}{\left(\sigma^2+i\frac{t}{m}\right)^{1\over 2}}
	e^{\mp i(\lambda+\epsilon x)t} e^{-i\frac{\epsilon^2}{6m}t^3}
	e^{ip_0x_0}e^{-{1\over 2}p_0^2\sigma^2}
	e^{-{1\over 2}\frac{\left(x-x_0\pm\frac{\epsilon}{2m}t^2-ip_0\sigma^2\right)^2}{\sigma^2+i\frac{t}{m}}}.
\end{equation}
We can easily see that we have got an entangled state between the measuring device made by the
magnet and the quantum system. This matter is all well-known in literature. Apparently there
is now way to get rid of the interference term arising from this entanglement while
Stern-Gerlach experiment gives rise to two clearly distinct traces for each spin component.
This aspect represents the main question of the measurement problem in quantum mechanics.

Bohm \cite{bohm} proposed a way out to this situation by observing that in the interference
terms there appear strongly oscillating in time phase factors and, wherever a good measurement
should be carried out, at different measurements the phase appear to be random. We can recognize
here, with a different formulation, the idea of a singular limit in time recently
proposed by us \cite{fra1,fra2,fra3,fra4} and pioneered by Bohm. The effectiveness of this
argument for the Stern-Gerlach apparatus is granted if the Bohm limit $\epsilon \Delta t /\Delta p\gg 1$
holds. We have set $\Delta t = l/v$ with $l$ the length of the magnet and $v$ the velocity of
the particle, while $\Delta p$ is the spreading in momentum. The argument due to Bohm do not
need any external environment.

Now, we observe that the Bohm argument take the system into a premeasurement status, that is, we are into
the situation proper to decoherence of a mixed form of the density matrix. If the Bohm limit
is violated, we could not be able to see the setup properly working. Studying the
transition between the two regimes is then possible and very interesting, taking us to the
limit of the ``decision point''\cite{har} into the transition region.

The above computation proves that our Stern-Gerlach device has proper eigenstates of spin
those of the $z$-axis. Then, assuming as initial eigenstates $\sigma_z|\uparrow\rangle=|\uparrow\rangle$ and
$\sigma_z|\downarrow\rangle=-|\downarrow\rangle$, we get the average values for position and
momentum
\begin{eqnarray}
   \langle x \rangle_\uparrow &=& x_0 + \frac{p_0}{m}t-{1\over 2}{\epsilon\over m} t^2 \\ \nonumber
   \langle x \rangle_\downarrow &=& x_0 + \frac{p_0}{m}t+{1\over 2}{\epsilon\over m} t^2 \\ \nonumber
   \langle p \rangle_\uparrow &=& p_0 - \epsilon t \\ \nonumber
   \langle p \rangle_\downarrow &=& p_0 + \epsilon t  
\end{eqnarray}
that, as it should be expected on the basis of the Ehrenfest theorem \cite{mess}, 
are the solutions of the classical Hamilton equations. In the same way, we have for 
quantum fluctuations
\begin{eqnarray}
   \langle x^2 \rangle_{\uparrow,\downarrow}-\langle x \rangle_{\uparrow,\downarrow}^2 &=&  
   {\sigma^2\over 2} + {t^2\over 2m^2\sigma^2} \\ \nonumber
   \langle p^2 \rangle_{\uparrow,\downarrow}-\langle p \rangle_{\uparrow,\downarrow}^2 &=&  
   p_0^2+{1\over 2\sigma^2}
\end{eqnarray}
and we see that just the spreading of the initial wave-packet, as the particle would be free,
is playing a role here. Now, checking the relative fluctuations one has
\begin{equation}
   \frac{\langle x^2 \rangle_{\uparrow,\downarrow}-\langle x \rangle_{\uparrow,\downarrow}^2}
   {\langle x \rangle_{\uparrow,\downarrow}^2}\approx \frac{2}{\sigma^2\epsilon^2t^2}
\end{equation}
in the limit of enough large times. We realize immediately that the only way to get a classical
behaviour for the particle is to have the Bohm limit to hold. Otherwise, the Stern-Gerlach
device will not work properly. As said above, this argument runs just if we consider eigenstates
of the z-component of the spin. 

So, turning back to a superposition state, one has still to understand the proper working of
the Stern-Gerlach device as a selector for the spin components. The Bohm argument and, generally
speaking, any argument relying on decoherence put us in the situation to have the density matrix
in the mixed form
\begin{equation}
    \rho_M(x,x',t) = |a|^2\psi_\uparrow^*(x,t)\psi_\uparrow(x',t)|\uparrow\rangle\langle\uparrow|
	+|b|^2\psi_\downarrow^*(x,t)\psi_\downarrow(x',t)|\downarrow\rangle\langle\downarrow|.
\end{equation}
Now, the diagonal part of this matrix has a very interesting property. That is, if one analyzes
the time evolution one can see that the possible initial overlapping of the two gaussians is
rapidly lost to mimic a ``collapsed'' wave function. That is, if the Bohm limit holds, then
the time evolution grants a rapid localization of the particle on the proper path and the
above mixed form is enough, in this case, to grant the proper working of the device. This is
due to the fast going away of the mean values of the gaussians one each other, making even more
improbable the possibility to observe intermediate results with respect to the two spin projections.

In conclusion, we have shown how the proper working of a Stern-Gerlach device can be understood
by the validity of the Bohm limit. Such a description, described by Bohm in \cite{bohm}, pioneered the
concept of a singular limit given in \cite{berry,fra1,fra2,fra3,fra4} to a proper understanding of
the classical limit. The possible verification of the behavior of the Stern-Gerlach device in the
transition region delimited by the Bohm limit should be explored experimentally as belongs to
the region where the ``decision'', intended as the region where a measurement device acts as
expected by Copenaghen interpretation, happens.

\end{document}